\documentclass[runningheads]{llncs}

\usepackage{amsmath}
\usepackage{mathtools}
\usepackage{epsfig}
\usepackage{color}
\usepackage{verbatim}
\usepackage{amsfonts}
\usepackage{url}

\def\BWT{\mbox{\rm L}}

\def\reverse#1{\hat{#1}}

\def\SA{\mbox{\rm {\sf SA}}}
\def\ISA{\mbox{\rm {\sf ISA}}}

\def\LF{\mbox{\rm {\sf LF}}}
\def\FMI{\mbox{\rm {\sf FMI}}}

\def\rank{\textsf{rank}}

\def\lcp{\mbox{\rm {\sf lcp}}}

\def\X{\textsf{X}}
\def\MS{\mbox{\rm {\sf MS}}}
\def\B{\textsf{B}}

\def\A{\textsf{A}}
\def\Pr{\mbox{\rm {\sf P}}}
\def\R{\mbox{\rm {\sf R}}}
\def\Y{\mbox{\rm {\sf Y}}}
\def\Y{\textsf{Y}}
\def\ssY{\mbox{\rm {\sf {\scriptsize Y}}}}
\def\Z{\textsf{Z}}
\def\C{\mbox{\rm {\sf C}}}

\def\BWT{\mbox{\rm {\sf BWT}}}
\def\LCP{\mbox{\rm {\sf LCP}}}
\def\LPF{\mbox{\rm {\sf LPF}}}

\def\YZ{\mbox{\rm {\sf {\scriptsize Y$|$Z}}}}

\def\NSV{\mbox{\rm {\sf NSV}}}
\def\PSV{\mbox{\rm {\sf PSV}}}
\def\RMQ{\mbox{\rm {\sf RMQ}}}
\def\LZSCAN{\mbox{\sf LZscan}}
\def\LZFMICN{\mbox{\sf LZ-FMI-CN}}
\def\LZFMIBPR{\mbox{\sf LZ-FMI-BPR}}
\def\LZFMIISA{\mbox{\sf LZ-FMI-ISA}}

\def\O{\mbox{\rm O}}

\def\balgorithm#1{{\bf Algorithm #1}}

\def\bfor{{\bf for\ }}

\def\bto{{\bf to\ }}
\def\bdownto{{\bf downto\ }}

\def\bdo{{\bf do\ }}
\def\bif{{\bf if\ }}
\def\bthen{{\bf then\ }}
\def\belse{{\bf else\ }}

\def\la{\leftarrow}

\def\+{\!+\!}
\def\-{\!-\!}

\begin{document}
\title{Lightweight Lempel-Ziv Parsing\thanks{Supported by Academy of Finland grant 118653 (ALGODAN)}}

\author{
Juha K{\"a}rkk{\"a}inen
\and
Dominik Kempa
\and
Simon J. Puglisi
}

\institute{
    Department of Computer Science,\\
    University of Helsinki\\
    Helsinki, Finland\\
    \email{\{firstname.lastname\}@cs.helsinki.fi}\\[1ex]
}

\date{}

\maketitle \thispagestyle{empty}

\begin{abstract}

  We introduce a new approach to LZ77 factorization that uses
  $\O(n/d)$ words of working space and $\O(dn)$ time for any $d\ge 1$
  (for polylogarithmic alphabet sizes). We also describe carefully
  engineered implementations of alternative approaches to lightweight 
  LZ77 factorization. Extensive experiments show that the
  new algorithm is superior in most cases, particularly at the lowest
  memory levels and for highly repetitive data. As a part of the
  algorithm, we describe new methods for computing matching statistics
  which may be of independent interest.

\end{abstract}

\section{Introduction}
\label{sec-intro}

The Lempel-Ziv factorization~\cite{ZL77}, also known as the LZ77
factorization, or LZ77 parsing, is a fundamental tool for compressing
data and string processing, and has recently become the basis for
several compressed full-text pattern matching
indexes~\cite{kn2011,ggknp2012}. These indexes are designed to
efficiently store and search massive, highly-repetitive data sets ---
such as web crawls, genome collections, and versioned code
repositories --- which are increasingly common~\cite{n2012}.

In traditional compression settings (for example the popular {\tt
  gzip} tool) LZ77 factorization is kept timely by factorizing
relative to only a small, recent window of data, or by breaking the
data up into blocks and factorizing each block separately. This
approach fails to capture widely spaced repetitions in the input, and
anyway, in many applications, including construction of the above
mentioned LZ77-based text indexes, whole-string LZ77 factorizations
are required.

The fastest LZ77 algorithms (see~\cite{kp2013,kkp2013}) use a lot of
space, at least $6n$ bytes for an input of $n$
symbols and often more. This prevents them from scaling to really large
inputs. Space-efficient algorithms are
desirable even on smaller inputs, as they place less burden on the
underlying system.

One approach to more space efficient LZ factorization is to use
compressed suffix arrays and succinct data structures~\cite{nm2007}.
Two proposals in this direction are due to Kreft and
Navarro~\cite{kn2010} and Ohlebusch and Gog~\cite{og2011}.
In this paper, we describe
carefully engineered implementations of these algorithms. 
We also propose a new,
space-efficient variant of the recent $\ISA$ 
family of algorithms~\cite{kp2013}. 
Most compressed index implementations are build from the uncompressed
suffix array (SA) which requires $4n$ bytes. Our implementations are
instead based on
the Burrows-Wheeler transform (BWT), constructed directly in about $2$--$2.5n$
bytes using the algorithm of Okanohara and Sadakane~\cite{os2009}.
There also exists two online algorithms based on compressed
indexes~\cite{os2008,s2012} but they are not competitive in practice
in the offline context.

The main contribution of this paper is a new algorithm to compute the
LZ77 factorization without ever constructing SA or BWT for the whole
input.  At a high-level, the algorithm divides the input up into
blocks, and processes each block in turn, by first computing a pattern
matching index for the block, then scanning the prefix of the input
prior to the block through the index to compute longest-matches, which
are then massaged into LZ77 factors.  For a string of length $n$ and
$\sigma$ distinct symbols, the algorithm uses $n\log\sigma + \O(n\log
n/d)$ bits of space, and $\O(d n t_{\rank})$ time, where $d$ is 
the number of blocks, and $t_{\rank}$
is the time complexity of the rank operation over sequences with
alphabet size $\sigma$ (see e.g.~\cite{bgnn2010}). The $n\log\sigma$
bits in the space bound is for the input string itself which is
treated as read-only.

Our implementation of the new algorithm does not, for the most part,
use compressed or succinct data structures. The goal is to optimize
speed rather than space in the data structures, because we can use the
parameter $d$ to control the tradeoff.
Our experiments demonstrate that this approach is in most cases
superior to algorithms using compressed indexes.

As a part of the new algorithm, we describe new techniques for
computing matching statistics~\cite{cl1994} that may be of independent
interest. In particular, we show how to invert matching statistics,
i.e., to compute the matching statistics of a string $\B$ w.r.t.~a
string $\A$ from the matching statistics of $\A$ w.r.t.~$\B$, which
saves a lot of space when $\A$ is much longer than $\B$.

All our implementations operate in main memory only and thus need at
least $n$ bytes just to hold the input. Reducing the memory
consumption further requires some use of external memory, a direction
largely unexplored in the literature so far.  We speculate
that the scanning, block oriented nature of the new algorithm will allow
efficient secondary memory implementations, but that study is left for
the future.

\section{Basic Notation and Algorithmic Machinery}
\label{sec-preliminaries}

\paragraph{Strings.}
Throughout we consider a string $\X = \X[1,n] = \X[1]\X[2]\ldots
\X[n]$ of $|\X| = n$ symbols drawn from the alphabet $[0,\sigma-1]$.
We assume $\X[n]$ is a special ``end of string'' symbol, \$, smaller than
all other symbols in the alphabet.
The reverse of $\X$ is denoted $\reverse{\X}$.  For $i=1,\ldots,n$ we
write $\X[i,n]$ to denote the {\em suffix} of $\X$ of length $n-i+1$,
that is $\X[i,n] = \X[i]\X[i+1]\ldots \X[n]$.  We will often refer to
suffix $\X[i,n]$ simply as ``suffix $i$''. Similarly, we write
$\X[1,i]$ to denote the {\em prefix} of $\X$ of length $i$.
$\X[i,j]$ is the {\em substring} $\X[i]\X[i+1]\ldots \X[j]$ of $\X$
that starts at position $i$ and ends at position $j$. By $\X[i,j)$ we
denote $\X[i,j-1]$.  If $j < i$ we
define $\X[i,j]$ to be the empty string, also denoted by
$\varepsilon$.

\paragraph{Suffix Arrays.}
The suffix array~\cite{mm1993} $\SA_{\X}$ (we drop subscripts when
they are clear
from the context) of a string $\X$ 
is an array $\SA[1,n]$ which
contains a permutation of the integers $[1,n]$ such that $\X[\SA[1],n]
< \X[\SA[2],n] < \cdots < \X[\SA[n],n]$.  In other words, $\SA[j] =
i$ iff $\X[i,n]$ is the $j^{\mbox{{\scriptsize th}}}$ suffix of $\X$
in ascending lexicographical order. The inverse
suffix array $\ISA$ is the inverse permutation of $\SA$, that is
$\ISA[i] = j$ iff $\SA[j] = i$.

Let $\lcp(i,j)$ denote the length of the longest-common-prefix of
suffix $i$ and suffix $j$. For example, in the string $\X =
zzzzzapzap$, $\lcp(1,4) = 2 = |zz|$, and $\lcp(5,8) = 3 = |zap|$. The
longest-common-prefix (LCP) array~\cite{klaap2001,KarkkainenMP09},
$\LCP_{\X}=\LCP[1,n]$, is defined such that $\LCP[1] = 0$, and $\LCP[i] =
\lcp(\SA[i],\SA[i-1])$ for $i \in [2,n]$.  
 
For a string $\Y$, the $\Y$-interval in the suffix array $\SA_{\X}$ is
the interval $\SA[s,e]$ that contains all suffixes having $\Y$ as a
prefix. The $\Y$-interval is a representation of the occurrences of
$\Y$ in $\X$. For a character $c$ and a string $\Y$, the computation
of $c\Y$-interval from $\Y$-interval is called a \emph{left extension}
and the computation of $\Y$-interval from ${\Y}c$-interval is called a
\emph{right contraction}. \emph{Left contraction} and \emph{right
  extension} are defined symmetrically.

\paragraph{BWT and backward search.}
The Burrows-Wheeler Transform~\cite{bw1994} $\BWT[1,n]$ is a
permutation of $\X$ such that $\BWT[i] = \X[\SA[i]-1]$ if $\SA[i]>1$
and $\$$ otherwise. We also define $\LF[i] = j$ iff $\SA[j] =
\SA[i]-1$, except when $\SA[i] = 1$, in which case $\LF[i] = \ISA[n]$.
Let $\C[c]$, for symbol $c$ be the number of symbols
in $\X$ lexicographically smaller than $c$.  The function
$\rank(\X,c,i)$, for string $\X$, symbol $c$, and integer $i$, returns
the number of occurrences of $c$ in $\X[1,i]$.  It is well known that
$\LF[i] = \C[\BWT[i]] + \rank(\BWT,\BWT[i],i)$.  Furthermore, we can
compute the left extension using $\C$ and $\rank$.  If $\SA[s,e]$ is
the $\Y$-interval,
then 
$\SA[\C[c]+\rank(\BWT,c,s),\C[c]+\rank(\BWT,c,e)]$ is 
the $c\Y$-interval.
This is called \emph{backward search}.

\paragraph{NSV/PSV and RMQ.} For an array $\A$, the \emph{next and previous
  smaller value} (NSV/PSV) operations are defined as
  $\NSV[i] = \min \{ j\in [i+1,n] \mid \A[j] < \A[i]\}$
and
  $\PSV[i] = \max \{ j\in [1,i-1] \mid \A[j] < \A[i]\}$.  
A related operation on $\A$ is \emph{range minimum
  query}: $\RMQ(\A,i,j)$ is $k\in[i,j]$ such that $\A[k]$ is the
minimum value in $\A[i,j]$. Both NSV/PSV operations and RMQ operations
over the LCP array can be used for implementing right contraction (see
Section~\ref{sec-ms}).

\paragraph{LZ77.}
Before defining the LZ77 factorization, we introduce the concept of a
{\em longest previous factor} (LPF).  The LPF at position $i$ in
string $\X$ is a pair $\LPF_{\X}[i]=(p_i,\ell_i)$ such that, $p_i < i$,
$\X[p_i,p_i+\ell_i) = \X[i,i+\ell_i)$, and $\ell_i$ is maximized.
In other words, $\X[i,i+\ell_i)$ is the longest
prefix of $\X[i,n]$ which also occurs at some position $p_i < i$ in
$\X$. Note that if $\X[i]$ is the leftmost occurrence of a symbol in
$\X$ then $p_i$ does not exist. In this case we adopt the convention
that $p_i = \X[i]$ and $\ell_i = 0$. When $p_i$ does exist we call
$\X[p_i,p_i+\ell_i)$ the {\em source} for position $i$. Note also
that there may be more than one potential source (that is, $p_i$
value), and we do not care which one is used.

The LZ77 factorization (or LZ77 parsing) of a string $\X$ is then just
a greedy, left-to-right parsing of $\X$ into longest previous
factors. More precisely, if the $j$th LZ factor (or {\em phrase}) in
the parsing is to start at position $i$, then we output $(p_i,\ell_i)$
(to represent the $j$th phrase), and then the $(j+1)$th phrase starts
at position $i+\ell_i$, unless $\ell_i = 0$, in which case the next
phrase starts at position $i+1$.  When $\ell_i > 0$, the substring
$\X[p_i,p_i+\ell_i)$ is called the {\em source} of phrase
$\X[i,i+\ell_i)$. We denote the number of phrases in the
LZ77 parsing of $\X$ by $z$.

\paragraph{Matching Statistics.}
Given two strings $\Y$ and $\Z$, the matching statistics of $\Y$
w.r.t.~$\Z$, denoted $\MS_{\YZ}$ is an array of $|\Y|$ pairs,
$(p_1,\ell_1)$, $(p_2,\ell_2)$, ..., $(p_{|\ssY|},\ell_{|\ssY|})$,
such that for all $i \in [1,|\Y|]$, $\Y[i,i+\ell_i) =
\Z[p_i,p_i+\ell_i)$ is the longest substring starting at position $i$
in $\Y$ that is also a substring of $\Z$. The observant reader will
note the resemblance to the LPF array. Indeed, if we replace
$\LPF_{\Y}$ with $\MS_{\YZ}$ in the computation of the LZ factorization
of $\Y$, the result is the relative LZ factorization of $\Y$
w.r.t.~$\Z$~\cite{RLZspire2010}.

\section{Lightweight, Scan-based LZ77 Parsing}
\label{sec-algorithm}

In this section we present a new algorithm for LZ77 factorization
called \LZSCAN.

\paragraph{Basic Algorithm.}

Conceptually \LZSCAN\ divides $\X$ up into $d=\lceil n/b \rceil$ fixed size
blocks of length $b$: $\X[1,b]$, $\X[b+1,2b]$, ... . The last
block could be smaller than $b$, but this does not change the
operation of the algorithm.  In the description that follows we will
refer to the block currently under consideration as $\B$, and to the
prefix of $\X$ that ends just before $\B$ as $\A$. Thus, if $\B =
\X[kb+1,(k+1)b]$, then $\A = \X[1,kb]$.  

To begin with we will assume no LZ factor or its source
crosses a boundary of the block $\B$. Later we will show how to remove
these assumptions.

The outline of the algorithm for processing a block $\B$ is shown below.
\begin{enumerate}
\item Compute $\MS_{\A|\B}$
\item Compute $\MS_{\B|\A}$ from $\MS_{\A|\B}$, $\SA_{\B}$ and $\LCP_{\B}$
\item Compute $\LPF_{\A\B}[kb+1,(k+1)b]$ from $\MS_{\B|\A}$ and
  $\LPF_{\B}$
\item Factorize $\B$ using $\LPF_{\A\B}[kb+1,(k+1)b]$
\end{enumerate}
Step~1 is the computational bottleneck of the algorithm in theory and
practice. Theoretically, the time complexity of Step~1 is
$\O((|A|+|B|)t_{\rank})$, where $t_{\rank}$ is the time complexity of
the rank operation on $\BWT_{\B}$ (see, e.g.,~\cite{bgnn2010}). 
Thus the total time complexity of
\LZSCAN\ is $\O(dnt_{\rank})$ using $\O(b)$ words of space in
addition to input and output. 
The practical implementation of Step~1 is described
in
Section~\ref{sec-ms}. In the rest of this section,
we describe the details of the other steps.

\paragraph{Step 2: Inverting Matching Statistics.}

We want to compute $\MS_{\B|\A}$ but we cannot afford the space of
the large data structures on $\A$ required by standard methods.
Instead, we compute first $\MS_{\A|\B}$ involving large data
structures on $\B$, which we can afford, and only a scan of $\A$ (see
Section~\ref{sec-ms} for details). We then \emph{invert} $\MS_{\A|\B}$ to
obtain $\MS_{\B|\A}$.
The inversion algorithm is given in Fig.~\ref{fig-msinvert}.

\begin{figure}
  \begin{tabbing}
    00: \=\qquad\=\qquad\=\qquad\=\qquad\=\qquad\=\kill
    \balgorithm{MS-Invert}\\
    1:\>\bfor $i \la 1$ \bto $|\B|$ \bdo $\MS_{\B|\A}[i] \la (0,0)$\\
    2:\>\bfor $i \la 1$ \bto $|A|$ \bdo\\
    3:\>\>$(p_{\A},\ell_{\A}) \la \MS_{\A|\B}[i]$\\
    4:\>\>$(p_{\B},\ell_{\B}) \la \MS_{\B|\A}[p_{\A}]$\\
    5:\>\>\bif $\ell_{\A} > \ell_{\B}$ \bthen $\MS_{\B|\A}[p_{\A}] \la (i,\ell_{\A})$\\
    6:\>$(p,\ell) \la \MS_{\B|\A}[\SA_{\B}[1]]$\\
    7:\>\bfor $i \la 2$ \bto $|\B|$ \bdo\\
    8:\>\>$\ell \la \min(\ell,\LCP_{\B}[i])$\\
    9:\>\>$(p_{\B},\ell_{\B}) \la \MS_{\B|\A}[\SA_{\B}[i]]$\\
    10:\>\>\bif $\ell > \ell_{\B}$ \bthen
$\MS_{\B|\A}[\SA_{\B}[i]] \la (p,\ell)$\\
    11:\>\>\belse
$(p,\ell) \la (p_{\B},\ell_{\B})$\\
    12:\>$(p,\ell) \la \MS_{\B|\A}[\SA_{\B}[|\B|]]$\\
    13:\>\bfor $i \la |B|-1$ \bdownto $1$ \bdo\\
    14:\>\>$\ell \la \min(\ell,\LCP_{\B}[i+1])$\\
    15:\>\>$(p_{\B},\ell_{\B}) \la \MS_{\B|\A}[\SA_{\B}[i]]$\\
    16:\>\>\bif $\ell > \ell_{\B}$ \bthen
$\MS_{\B|\A}[\SA_{\B}[i]] \la (p,\ell)$\\
    17:\>\>\belse
$(p,\ell) \la (p_{\B},\ell_{\B})$
  \end{tabbing}
  \caption{Inverting matching statistics}
  \label{fig-msinvert}
\end{figure}

Note that the algorithm accesses each entry of $\MS_{\A|\B}$ only once
and the order of these accesses does not matter. Thus we can execute
the code on lines~3--5 immediately after computing $\MS_{\A|\B}[i]$
in Step~1 and then discard that value. This way we can avoid storing
$\MS_{\A|\B}$.

\paragraph{Step3: Computing LPF.}

Consider the pair $(p,\ell)=\LPF_{\A\B}[i]$ for $i\in[kb+1,(k+1)b]$
that we want to compute and assume $\ell>0$ (otherwise $i$ is the position
of the leftmost occurrence of $\X[i]$ in $\X$, which we can easily detect).
Clearly, either $p\le kb$ and $\LPF_{\A\B}[i]=\MS_{\B|\A}[i]$, or $kb < p < i$
and $\LPF_{\A\B}[i]=(kb+p_{\B},l_{\B})$, where $(p_{\B},l_{\B})=\LPF_{\B}[i-kb]$.
Thus computing $\LPF_{\A\B}$ from
$\MS_{\B|\A}[i]$ and $\LPF_{\B}$ is easy. 

The above is true if the sources do not cross the block
boundary, but the case where $p\le kb$ but $p+\ell > kb+1$ is not
handled correctly. An easy correction is to replace $\MS_{\A|\B}$ with
$\MS_{\A\B|\B}[1,kb]$ in all of the steps. 

\paragraph{Step 4: Parsing.}

We use the standard LZ77 parsing to factorize $\B$ except
$\LPF_{\B}$ is replaced with $\LPF_{\A\B}[kb+1,(k+1)b]$.

So far we have assumed that every block starts with a new phrase, or,
put another way, that a phrase ends at the end of every block. Let
$\X[i,(k+1)b]$ the last factor in $\B$, after we have factorized $\B$
as described above. This may not be a true LZ factor when considering
the whole $\X$ but may continue beyond the end of $\B$.
To find
the true end point, we treat
$\X[i,n]$ as a pattern, and apply the constant
extra space pattern matching algorithm of Crochemore~\cite{c1992}, looking for
the longest prefix of $\X[i,n]$ starting in $\X[1,i-1]$. We must
modify the algorithm from~\cite{c1992} so that it matches prefixes
rather than whole occurrences of the pattern, but this is possible
without increasing its time or space
complexity.

\section{Computation of matching statistics}
\label{sec-ms}

In this section, we describe how to compute the matching statistics 
$\MS_{\A|\B}$. As mentioned in Section~\ref{sec-algorithm}, what 
we really want is $\MS_{\A\B|\B}[1,kb]$. However, the only difference
is that the starting point of the computation is the $\B$-interval in
$\SA_{\B}$ instead of the $\varepsilon$-interval.

Similarly to most algorithms for computing the matching statistics, 
we first construct some data structures on $\B$ and then scan
$\A$. During the whole LZ factorization, most of the time is spend on
the scanning and the time for constructing the data structures is
insignificant in practice. Thus we omit the construction details here.
The space requirement of the data structures is more important but not
critical as we can compensate for increased space by
reducing the block size $b$. Using more space (per character of $\B$)
is worth doing if it increases scanning speed more than it increases
space. Consequently, we mostly use plain, uncompressed
arrays.

\paragraph{Standard  approach.}

The standard approach of computing the matching statistics using the
suffix array is to compute for each position $i$ the longest prefix
$\Pr_i=\A[i,i+\ell_i)$ of the suffix $\A[i,|A|]$ such that the
$\Pr_i$-interval in $\SA_{\B}$ is non-empty. Then
$\MS_{\A|\B}[i]=(p_i,\ell_i)$, where $p_i$ is any suffix in the
$\Pr_i$-interval.  This can be done either with a forward scan of
$\A$, computing each $\Pr_i$-interval from $\Pr_{i-1}$-interval using
the extend right and contract left operations~\cite{ako2004}, or with
a backward scan computing each $\Pr_i$-interval from
$\Pr_{i+1}$-interval using the extend left and contract right
operations~\cite{og2011}.  We use the latter alternative but with
bigger and faster data structures.

The extend left operation is implemented by backward search. We need
the array $\C$ of size $\sigma$ and an implementation of the rank
function on $\BWT$. For the latter, we use the fast
rank data structure of~\cite{fgm2012}, which uses $4b$ bytes.

The contract right operation is implemented using the NSV and PSV
operations on $\LCP_{\B}$ as in~\cite{og2011}, but instead of a
compressed representation, we store the NSV and PSV values as plain
arrays. As a nod towards reducing space, we store the NSV/PSV values
as offsets using 2 bytes each. If the offset is too large (which is
very rare), we obtain the value using the NSV/PSV data structure of
C{\'a}novas and Navarro~\cite{cn2010}, which needs less than $0.1b$ bytes.
Here the space saving was worth it as it had essentially no effect on
speed.

The peak memory use of the resulting algorithm is
$n+(24.1)b+\O(\sigma)$ bytes.

\paragraph{New approach.}

Our second approach is similar to the first, but instead of
maintaining both end points of the $\Pr_i$-interval, we keep just one,
arbitrary position $s_i$ within the interval. In principle, we perform
left extension by backward search, i.e.,
$s_i=\C[\X[i]]+\rank(\BWT,\X[i],s_{i+1})$. However, checking whether
the resulting interval is empty and performing right contractions if
it is, is more involved. To compute $s_i$ and $\ell_i$ from
$s_{i+1}$ and $\ell_{i+1}$, we execute the following steps:
\begin{enumerate}
\item Let $c=\X[i]$. If $\BWT[s_{i+1}]=c$, set
  $s_i=\C[c]+\rank(\BWT,c,s_{i+1})$ and $\ell_i=\ell_{i+1}+1$. 
\item Otherwise, let $\BWT[u]$ be the nearest occurrence of $c$ in
  $\BWT$ before the position $s_{i+1}$. Compute the rank of that
  occurrence $r=\rank(\BWT,c,u)$ and $\ell_u=\LCP[\RMQ(\LCP,u+1,s_{i+1})]$. If
  $\ell_u\ge \ell_{i+1}$, set $s_i=\C[c]+r$ and
  $\ell_i=\ell_{i+1}+1$. 
\item Otherwise, let $\BWT[v]$ be the nearest occurrence of $c$ in
  $\BWT$ after the position $s_{i+1}$ and compute
  $\ell_v=\LCP[\RMQ(\LCP,s_{i+1}+1,v)]$. If $\ell_v \le \ell_u$, set
  $s_i=\C[c]+r$ and $\ell_i=\ell_u+1$.
\item Otherwise, set $s_i=\C[c]+r+1$ and $\ell_i=\min(\ell_{i+1},\ell_v)+1$.
\end{enumerate}

The implementation of the above algorithm is based on the arrays
$\BWT$, $\LCP$ and $\R[1,b]$, where $\R[i]=\rank(\BWT,\BWT[i],i)$.
All the above operations can be performed by scanning $\BWT$ and
$\LCP$ starting from the position $s_{i+1}$ and accessing one value in
$\R$. To avoid long scans, we divide $\BWT$ and $\LCP$ into blocks of
size $2\sigma$, and store for each block and each symbol $c$, the
values $r$, $\ell_u$ and $\ell_v$ that would get computed if scans
starting inside the block continued beyond the block boundaries.

The peak memory use is $n+27b+\O(\sigma)$
bytes. This is more than in the first approach, but this is more than
compensated by increased scanning speed.

\paragraph{Skipping repetitions.}

During the preceding stages of the LZ factorization, we have 
built up knowledge of repetition present in $\A$, which can be
exploited to skip (sometimes large) parts of $\A$ during the
matching-statistics scan.
Consider an LZ factor $\A[i,i+\ell)$. 
Because, by definition, $\A[i,i+\ell)$ occurs earlier in $\A$ too, any
source of an LZ factor of $\B$ that is completely inside $\A[i,i+\ell)$
could be replaced with an equivalent source in that earlier
occurrence.  Thus such factors can be skipped during the computation
of $\MS_{\A|\B}$ without an effect on the factorization.

More precisely, if during the scan we compute $\MS_{\A|\B}[j]=(p,k)$
and find that $i \leq j < j + k \leq i+\ell$ 
for an LZ factor $\A[i,i+\ell)$, we will compute $\MS_{\A|\B}[i-1]$
and continue the scanning from $i-1$. However, we will do this only
for long phrases with $\ell\ge 40$.  To compute $\MS_{\A|\B}[i-1]$
from scratch, we use right extension operations implemented by a
binary search on $\SA$.

To implement this ``skipping trick'' we use a bitvector of $n$ bits to
mark LZ77 phrase boundaries adding $0.125n$ bytes to the peak memory.

\section{Algorithms Based on Compressed Indexes}
\label{sec-oldalgs}

We went to some effort to ensure the baseline system used to evaluate
$\LZSCAN$ in our experiments was not a ``straw man''. This required
careful study and improvement of some existing approaches, which we
now describe.

\paragraph{FM-Index.} The main data structure in all the algorithms
below is an
implementation of the FM-index (FMI)~\cite{fm2005}. It consists of two
main components:
\begin{itemize}
\item \textit{$\BWT_{\X}$ with support for the rank operation.} This enables
  backward search and the LF operation as described in
  Section~\ref{sec-preliminaries}. We have tried several rank data
  structures and found the one by
  Navarro~\cite[Sect.~7.1]{nav2004} to be the best in practice.
\item \textit{A sampling of $\SA_{\X}$.} This together with the LF
  operation enables arbitrary $\SA$ access since
  $\SA[i]=\SA[\LF^k[i]]+k$ for any $k<\SA[i]$.
  The sampling rate is a major space--time tradeoff parameter.
\end{itemize}
In many implementations of FMI, the construction starts with computing
the uncompressed suffix array but we cannot afford the space.
Instead, we construct $\BWT$ directly using the algorithm of
Okanohara and Sadakane~\cite{os2009}.
The method uses
roughly $2$--$2.5n$ bytes of space
but destroys the text, which is required later during LZ
parsing. Thus, once we have $\BWT$, we build a rank structure over it
and use it to invert the $\BWT$. During the inversion process we
recover and store the text and gather the $\SA$ sample values.

\paragraph{CPS2 simulation.} The CPS2 algorithm~\cite{cps2008} is an
LZ parsing algorithm based on $\SA_{\X}$. To compute the LZ factor
starting at $i$, it computes the $\X[i,i+\ell)$-interval for
$\ell=1,2,3,\ldots$ as long as the $\X[i,i+\ell)$-interval contains a
value $p<i$, indicating an occurrence of $\X[i,i+\ell)$ starting at $p$.

The key operations in CPS2 are right extension and checking
whether an $\SA$ interval contains a value smaller than $i$.  Kreft
and Navarro~\cite{kn2010} as well as Ohlebusch and Gog~\cite{og2011}
are using $\FMI$ for $\reverse{\X}$, the reverse of $\X$, 
which allows simulating right extension
on $\SA_{\X}$ by left extension on $\SA_{\reverse{\X}}$. The
two algorithms differ in the way they implement the interval checks:
\begin{itemize}
\item Kreft and Navarro use the RMQ operation. They use the RMQ data
  structure by Fischer and Heun~\cite{fh2007} but we use the one by
  C{\'a}novas and Navarro~\cite{cn2010}. The latter is easy and fast to
  construct during BWT inversion but queries are slow without an
  explicit $\SA$. We speed up queries by replacing a general RMQ with
  the check whether the interval contains a value smaller than $i$.
  This implementation is called $\LZFMICN$.
\item Ohlebusch and Gog use NSV/PSV queries. The position $s$ of $i$
  in $\SA$ must be in the $\X[i,i+\ell)$-interval. Thus we just need
  to check whether either $\NSV[s]$ or $\PSV[s]$ is in the interval
  too. They as well as we implement NSV/PSV using a balanced
  parentheses representation (BPR). This representation is initialized
  by accessing the values of $\SA$ left-to-right, which makes the
  construction slow using $\FMI$. However, NSV/PSV queries with this data
  structure are fast, as they do not require accessing $\SA$.
  This implementation is called $\LZFMIBPR$.
\end{itemize}

\paragraph{ISA variant.} Among the most space efficient prior LZ
factorization algorithms are those of the ISA
family~\cite{kp2013} that use a sampled $\ISA$, a full $\SA$ and a
rank/LF implementation that relies on the presence of the full
$\SA$. We reduce the space further by replacing $\SA$ and the rank/LF
data structure with the FM-index described above to obtain an
algorithm called $\LZFMIISA$.

\clearpage

\section{Experiments}
\label{sec-experiments}

We performed experiments with 
the files listed in Table~\ref{tab:files}.  All tests were conducted
on a 2.53GHz Intel Xeon Duo CPU with 32GB main memory and 8192K L2
Cache. The machine had no other significant CPU tasks running. The
operating system was Linux (Ubuntu 10.04) running kernel 3.0.0-26. The
compiler was g++ (gcc version 4.4.3) executed with the {\tt -O3
  -static -DNDEBUG} options. Times were recorded with the C {\tt
  clock} function. All algorithms operate strictly in-memory.

\begin{figure}
\minipage{0.5\textwidth}
  \includegraphics[trim = 0mm 25mm 0mm 0mm, width=\linewidth]{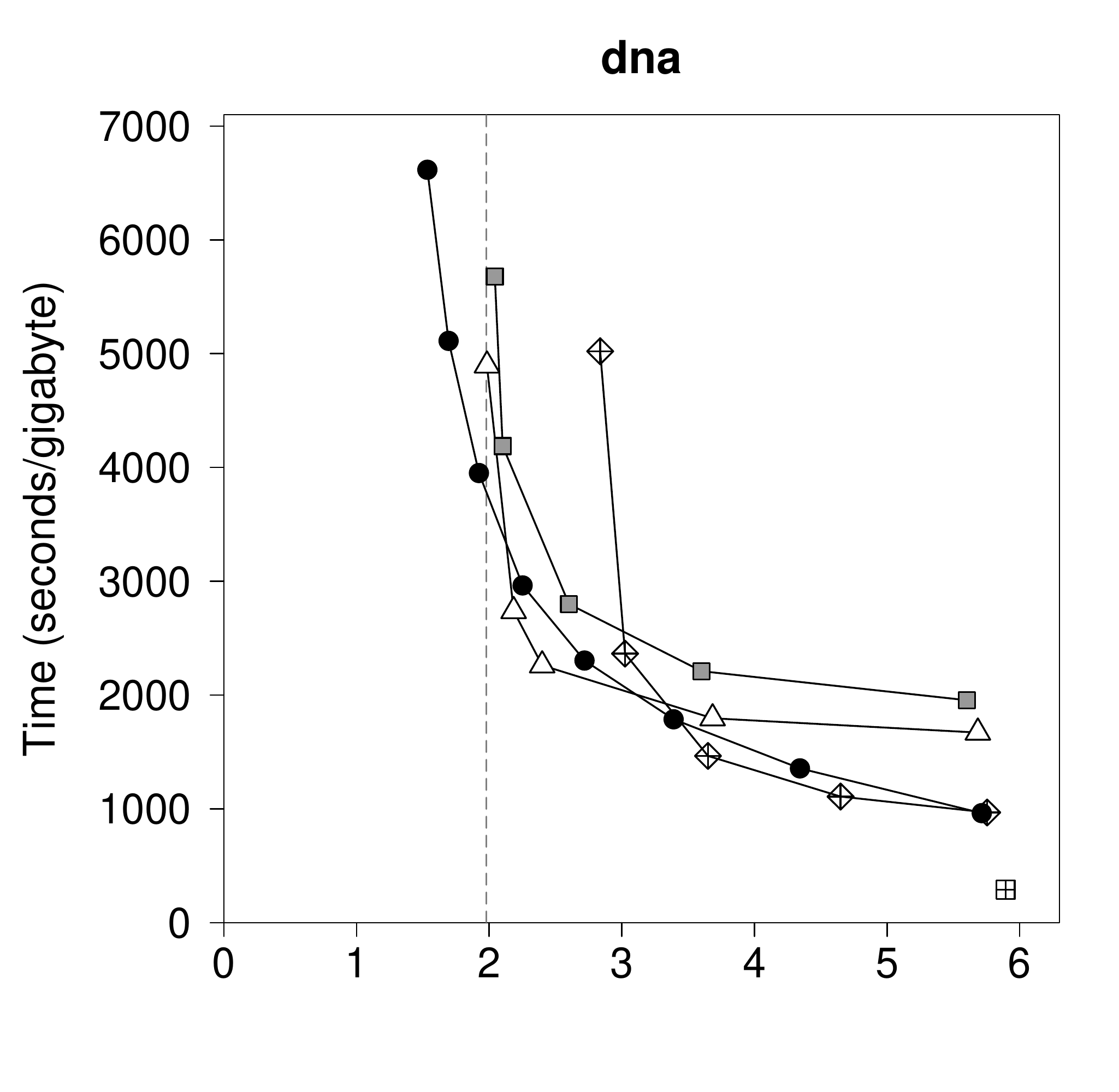}
\endminipage\hfill
\minipage{0.5\textwidth}
  \includegraphics[trim = 20mm 25mm -20mm 0mm, width=\linewidth]{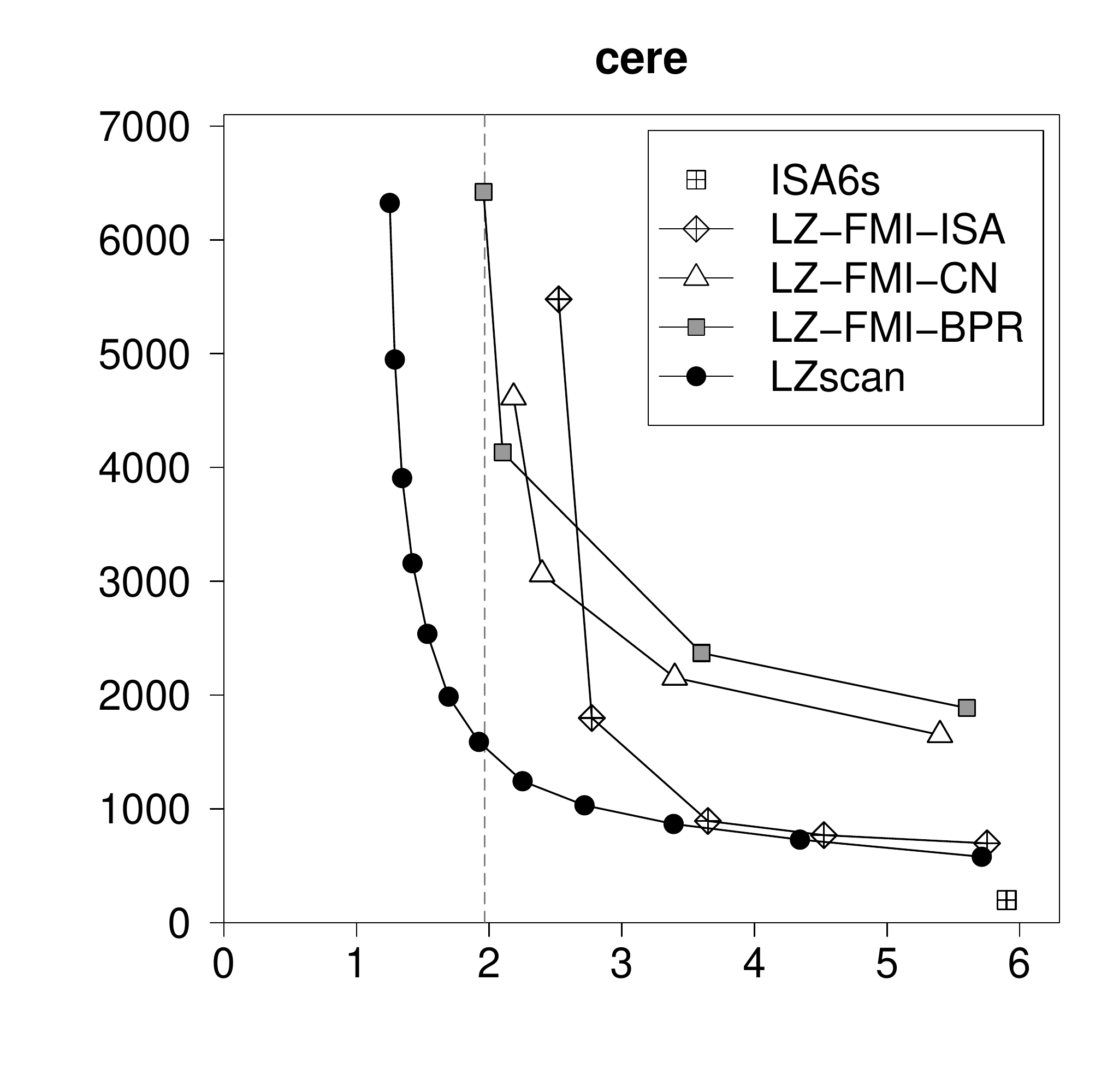}
\endminipage
\vspace{1ex}
\newline
\minipage{0.5\textwidth}
  \includegraphics[trim = 0mm 25mm 0mm 0mm, width=\linewidth]{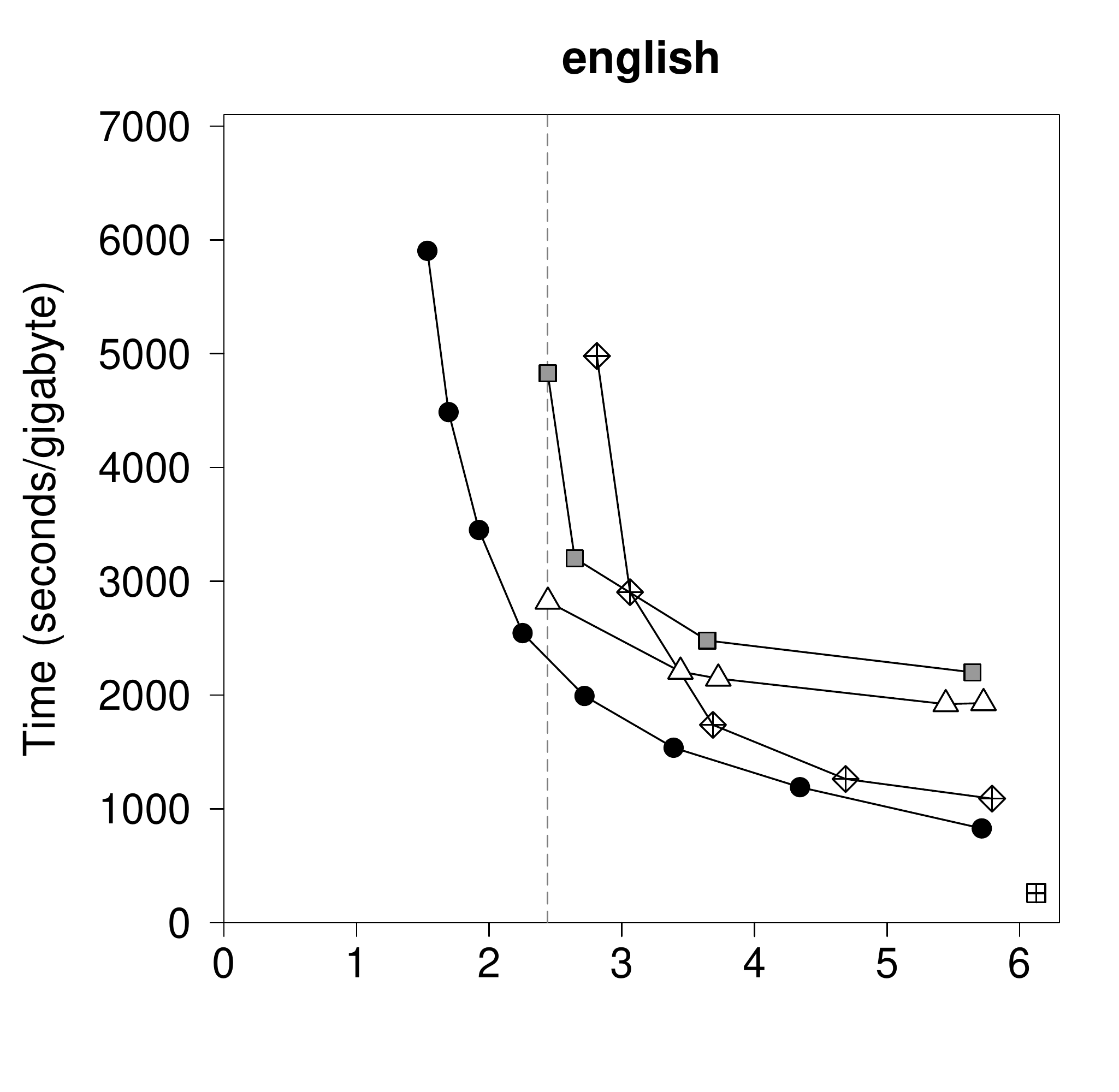}
\endminipage\hfill
\minipage{0.5\textwidth}
  \includegraphics[trim = 20mm 25mm -20mm 0mm, width=\linewidth]{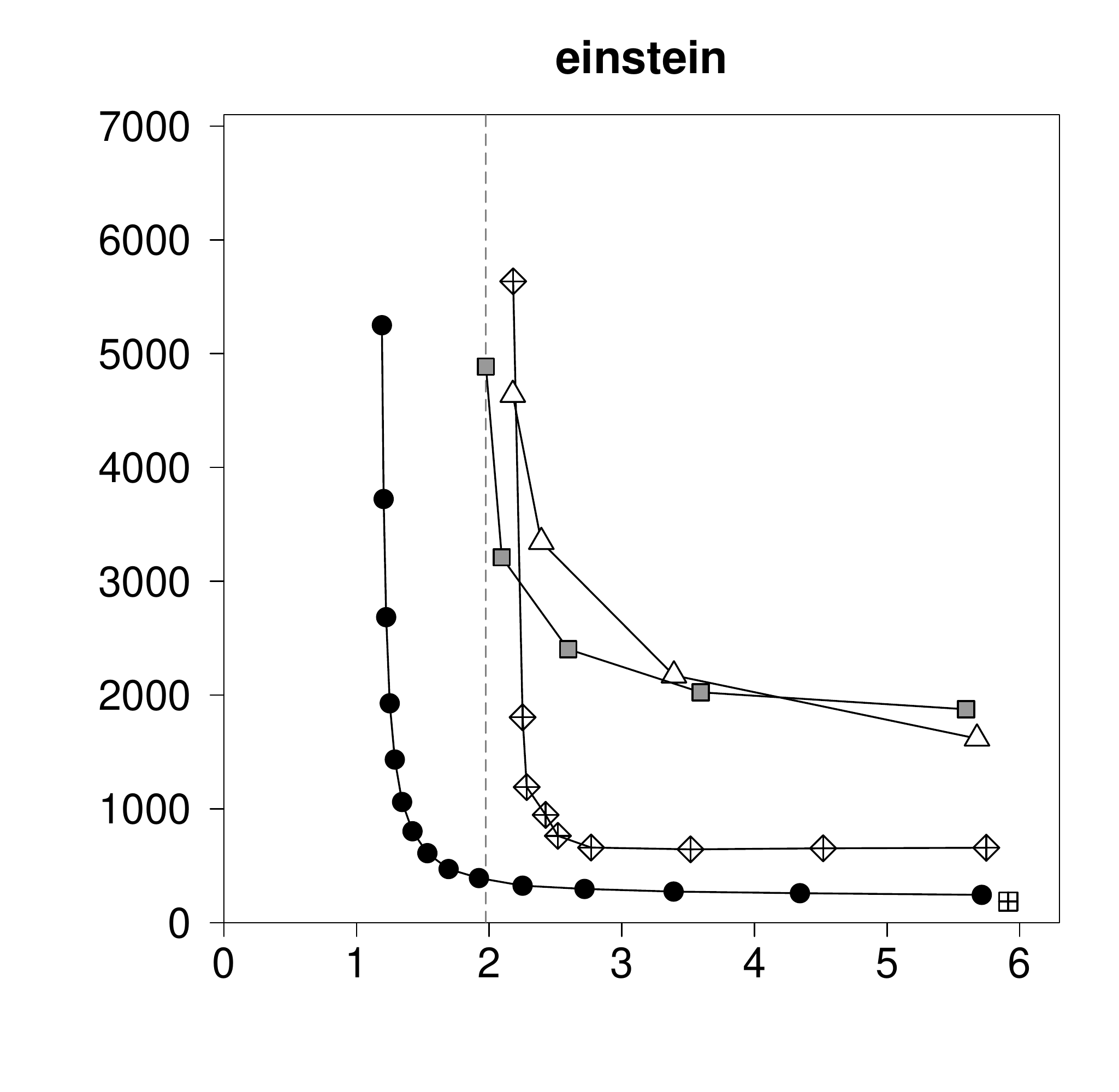}
\endminipage
\vspace{1ex}
\newline
\minipage{0.5\textwidth}
  \includegraphics[trim = 0mm 25mm 0mm 0mm, width=\linewidth]{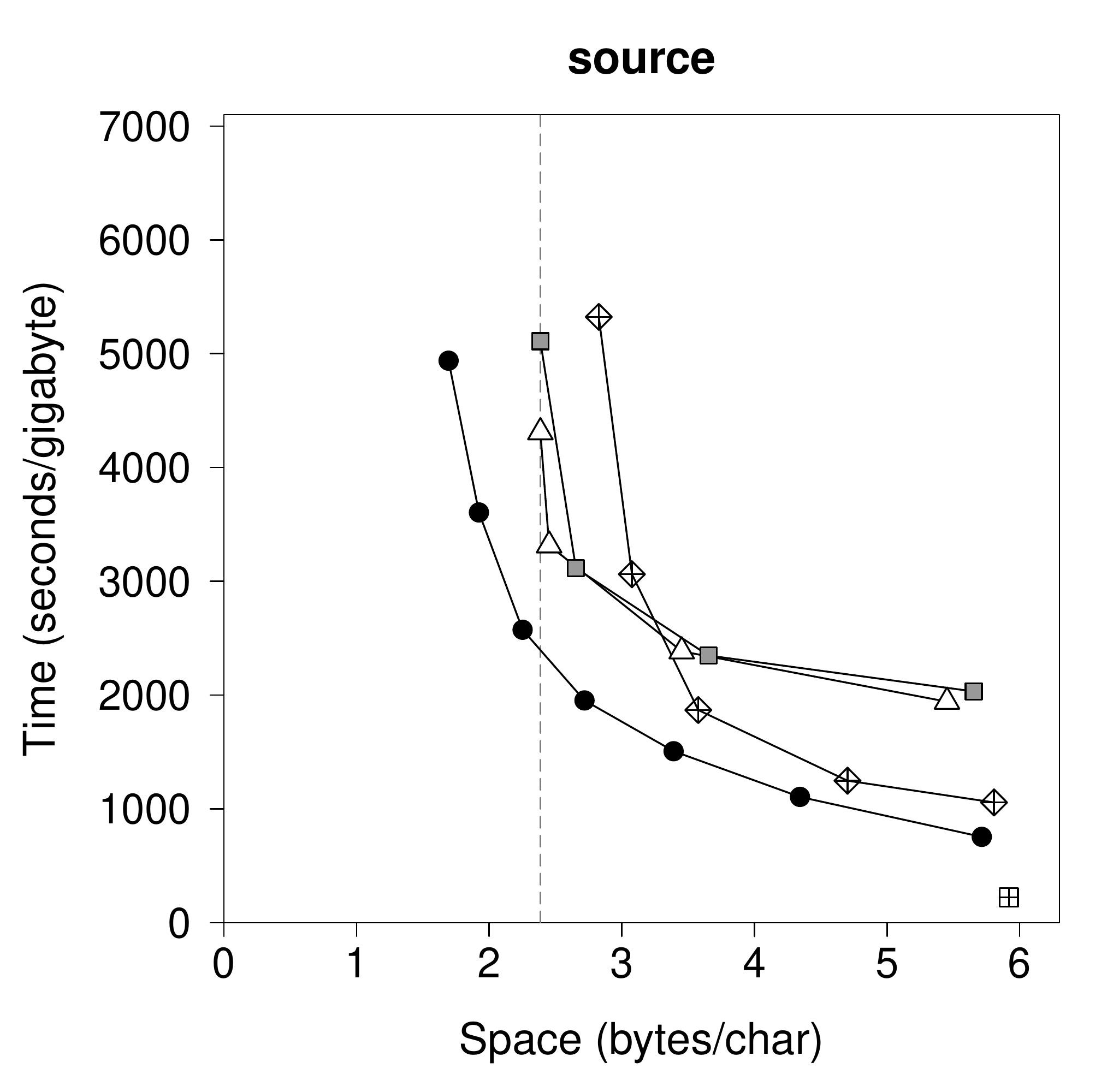}
\endminipage\hfill
\minipage{0.5\textwidth}
  \includegraphics[trim = 20mm 25mm -20mm 0mm, width=\linewidth]{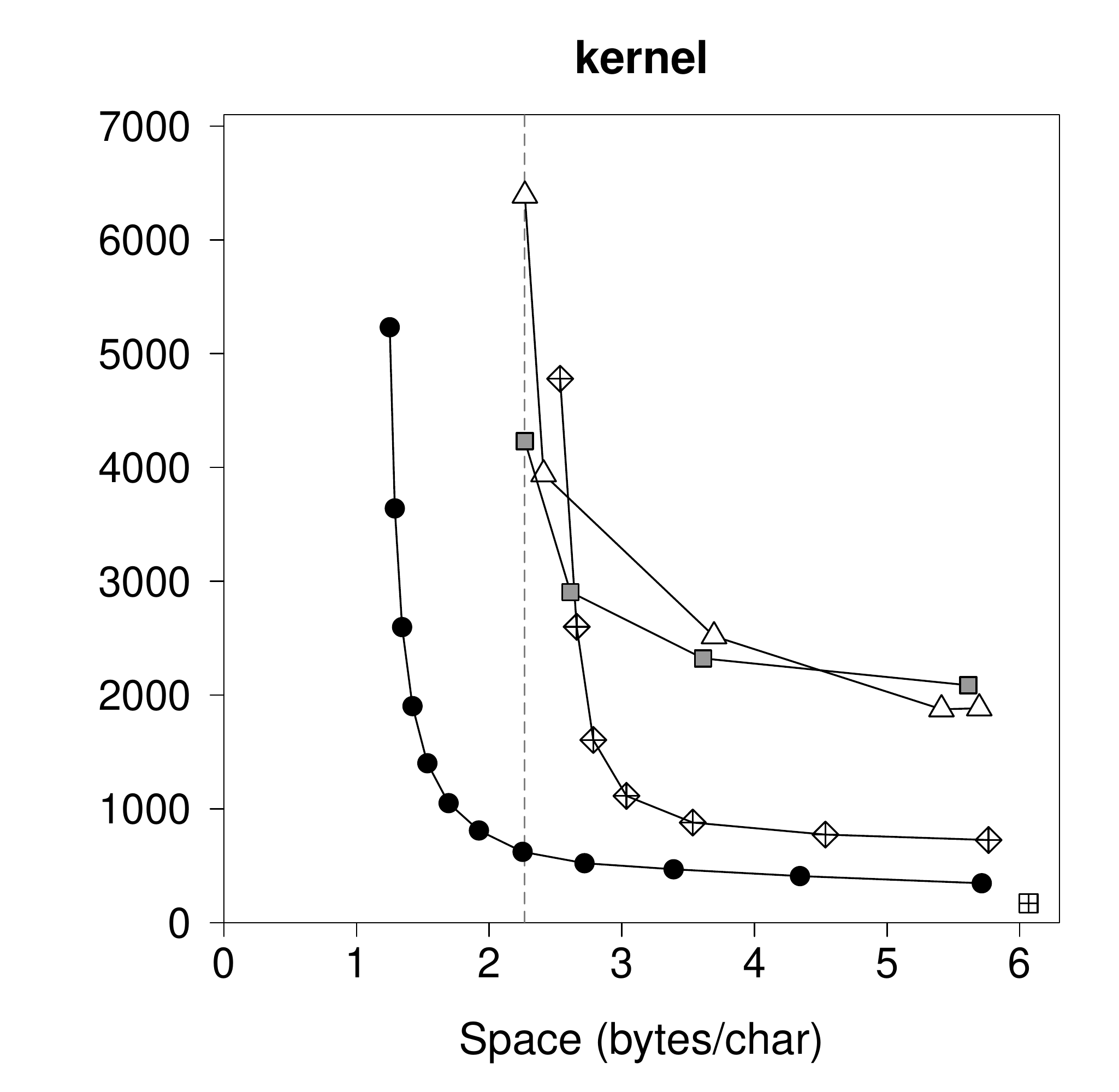}
\endminipage
\vspace{4ex}
\caption{\label{fig-all} Time-space tradeoffs for various LZ77
  factorization algorithms. The times do not include reading from or
  writing to disk. For algorithms with multiple parameters controlling
  time/space we show only the optimal points, that is, points forming
  the lower convex hull of the points ``cloud'' corresponding to various
  settings. The vertical line is the peak memory usage of $\BWT$
  construction algorithm \cite{os2009}. For comparison, we show the
  runtimes of $\ISA 6s$~\cite{kp2013}, currently
  the fastest LZ77 factorization algorithm using  $6n$ bytes.}
\end{figure}

\paragraph{LZscan vs. other algorithms.} We compared the $\LZSCAN$
implementation using our new approach for matching statistics boosted
with the ``skipping trick'' (Section~\ref{sec-ms}) to algorithms based
on compressed indexes (Section~\ref{sec-oldalgs}). The experiments measured
the time to compute LZ factorization with varying amount of available
working space. The results are shown in Figure~\ref{fig-all}. In almost
all cases $\LZSCAN$ outperforms other algorithm across the whole space
spectrum. Moreover, it can operate with very small available memory (close
to $n$ bytes) unlike other algorithms, which all require at least $2n$ space
to compute $\BWT$. It achieves a superior performance for highly
repetitive data even at very low memory levels.

\begin{figure}
\minipage{0.5\textwidth}
  \includegraphics[trim = 0mm 25mm 0mm 0mm, width=\linewidth]{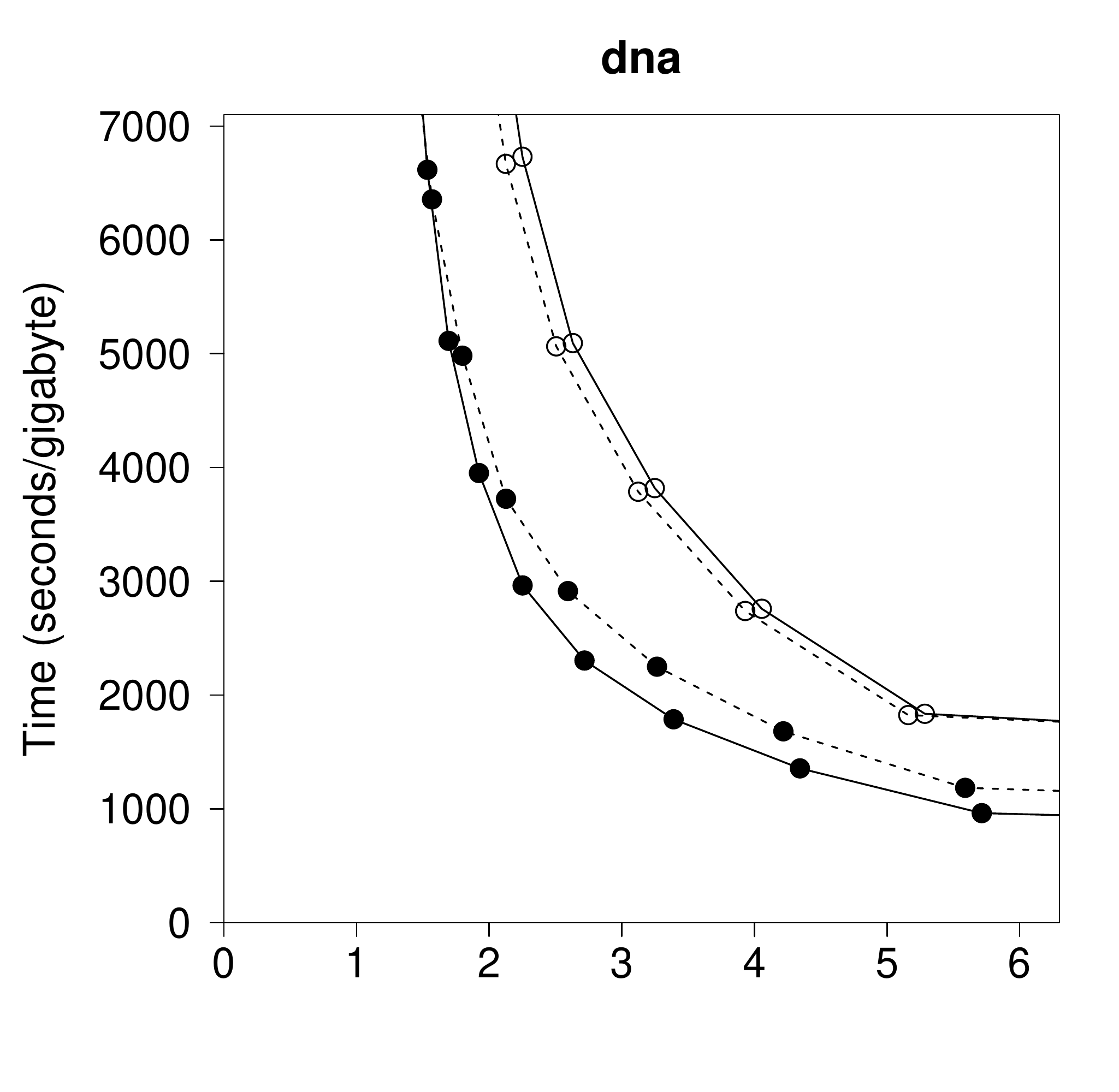}
\endminipage\hfill
\minipage{0.5\textwidth}
  \includegraphics[trim = 20mm 25mm -20mm 0mm, width=\linewidth]{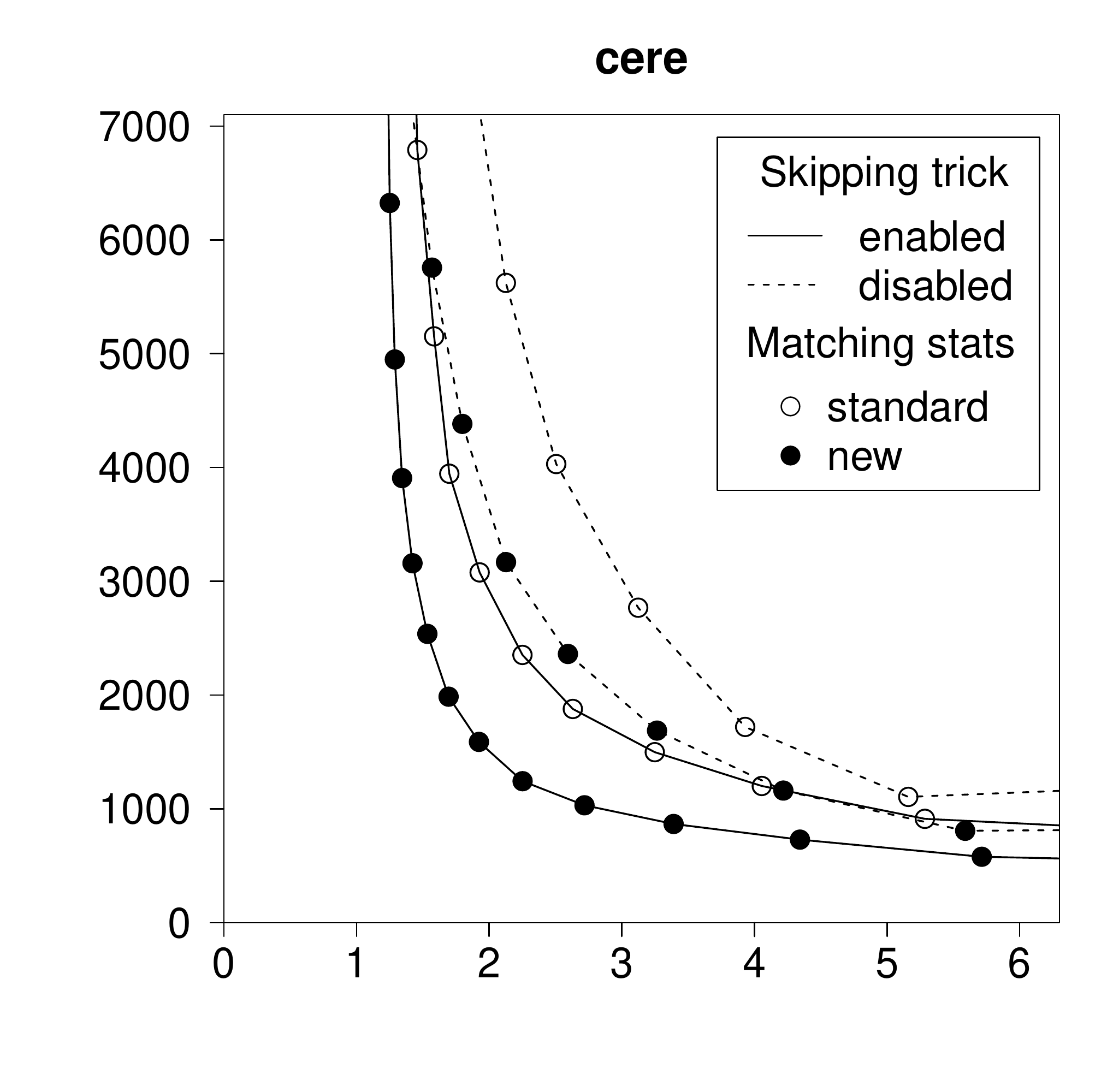}
\endminipage
\vspace{1ex}
\newline
\minipage{0.5\textwidth}
  \includegraphics[trim = 0mm 25mm 0mm 0mm, width=\linewidth]{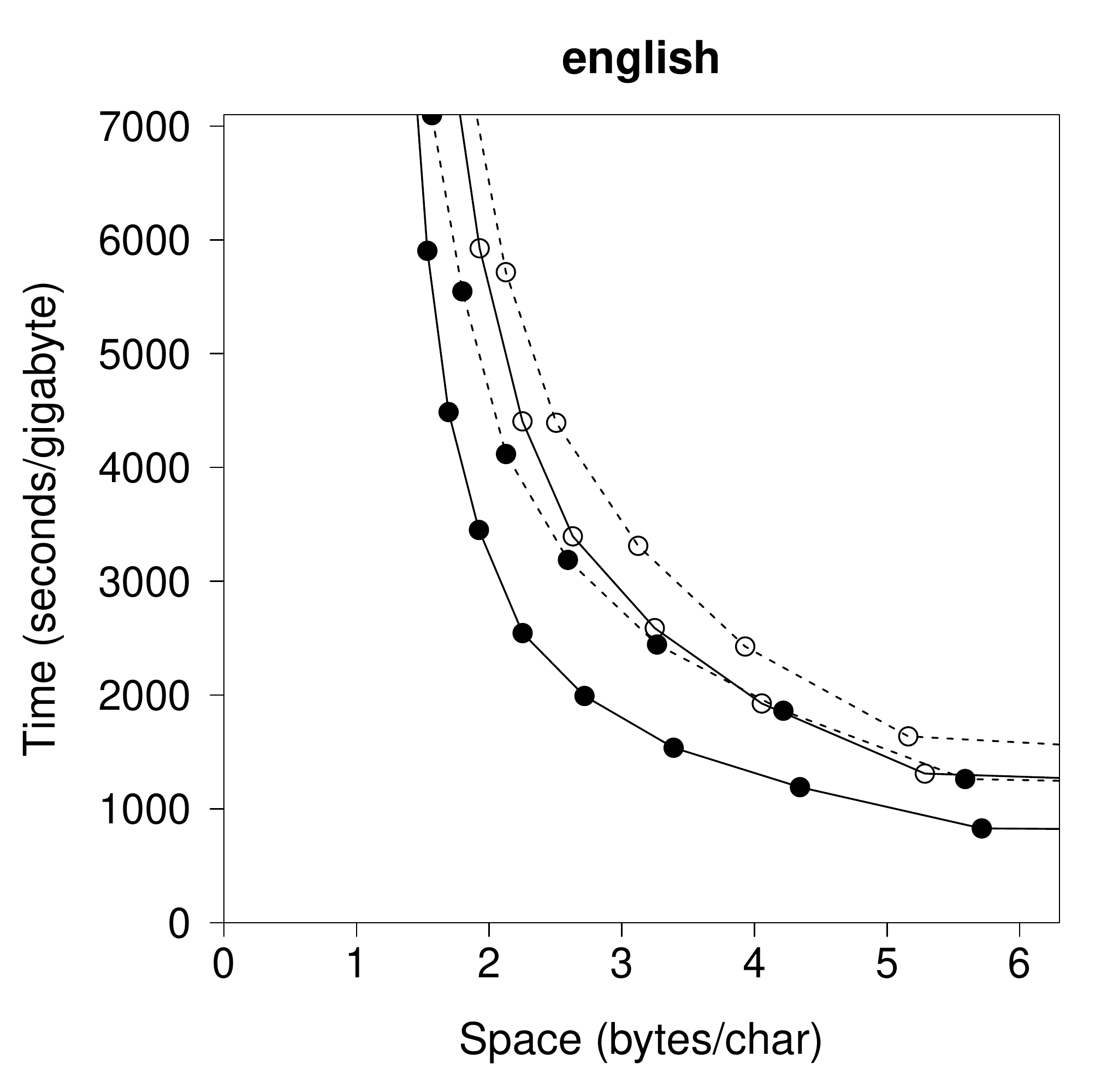}
\endminipage\hfill
\minipage{0.5\textwidth}
  \includegraphics[trim = 20mm 25mm -20mm 0mm, width=\linewidth]{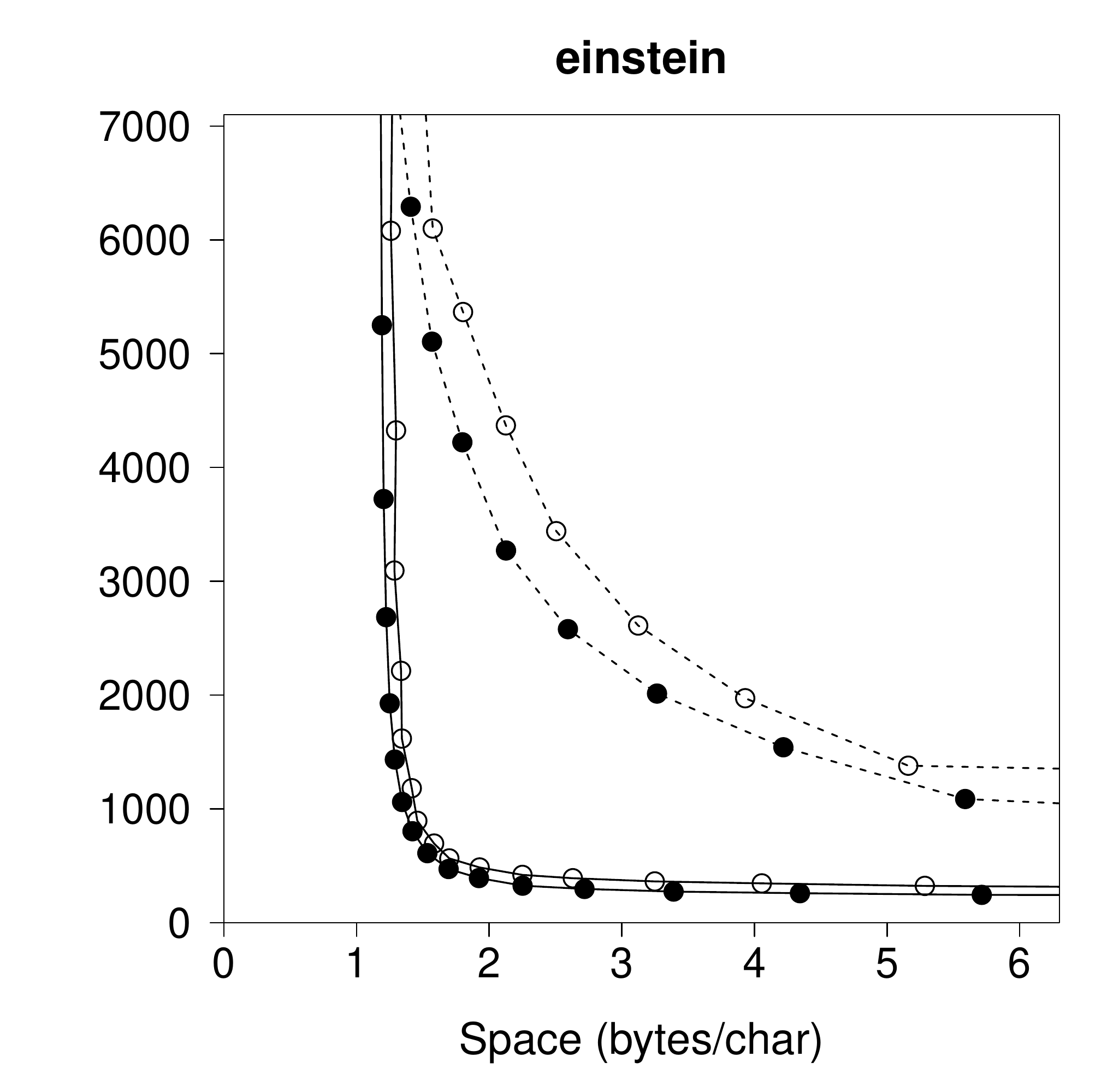}
\endminipage
\vspace{4ex}
\caption{\label{fig-variants} Time-space tradeoffs for different variants of
  $\LZSCAN$ algorithm. The variants differ in the subprocedure
  computing matching statistics. Each of two approaches can be additionally boosted
  by enabling the ``skipping trick'' yielding 4 different combinations.
  See Section~\ref{sec-ms} for details. The times do not include reading
  from or writing to disk.}
\end{figure}

\paragraph{Variants of LZscan.}
The second experiment measured the improvement of our new matching
statistics computation over standard approach (see Section~\ref{sec-ms}).
Additionally, each variant was tested with and without the ``skipping trick'',
giving 4 combinations in total. The results are plotted in
Figure~\ref{fig-variants}. In nearly all cases applying any of our new
techniques improves the runtime over the standard approach, but the best
effect in all cases is achieved when the techniques are combined together.
The total speedup then varies from a factor of 2 (dna) up to 12 (einstein),
clearly depending on the repetitiveness of input.

\begin{table}[bt]
  \centering {\small
  \begin{tabular}[tab:space-basic]{l@{\hspace{1em}}r@{\hspace{1em}}r@{\hspace{1em}}c@{\hspace{1em}}l@{\hspace{1em}}l}
\hline
Name & $\sigma$ & $n/z$ & $n/2^{20^{\vphantom{a}}}$ & Source & Description \\
\hline
dna & 16 & 14.2 & 100 & S & Human genome\\
english & 215 & 14.1 & 100 & S & Gutenberg Project \\
sources & 227 & 16.8 & 100 & S & Linux and GCC sources \\
\hline
cere & 5 & 84 & 100 & R & yeast genome\\
einstein & 121 & 2947 & 100 & R & Wikipedia articles\\
kernel & 160 & 156 & 100 & R & Linux Kernel sources\\
\hline
  \end{tabular} }\vspace{1ex}
\caption[Caption for LOF]
{Data set used in the experiments. The files are 100MB prefixes of files from the
  Pizza \& Chili  standard corpus\footnotemark[2] (S)  and the Pizza \& Chili
  repetitive corpus\footnotemark[3] (R).
  The value of $n/z$ (average length of an LZ77 phrase) is included as a
  measure of repetitiveness. }
  \label{tab:files}
\end{table}

\footnotetext[2]{\url{http://pizzachili.dcc.uchile.cl/texts.html}}
\footnotetext[3]{\url{http://pizzachili.dcc.uchile.cl/repcorpus.html}}

\bibliographystyle{abbrv}
\bibliography{lz}

\begin{thebibliography}{10}

\bibitem{ako2004}
M.~I. Abouelhoda, S.~Kurtz, and E.~Ohlebusch.
\newblock Replacing suffix trees with enhanced suffix arrays.
\newblock {\em Journal of Discrete Algorithms}, 2(1):53--86, 2004.

\bibitem{bgnn2010}
J.~Barbay, T.~Gagie, G.~Navarro, and Y.~Nekrich.
\newblock Alphabet partitioning for compressed rank/select and applications.
\newblock In {\em Proc. ISAAC}, LNCS 6507, pages 315--326, 2010.

\bibitem{bw1994}
M.~Burrows and D.~Wheeler.
\newblock A block sorting lossless data compression algorithm.
\newblock Technical Report 124, Digital Equipment Corporation, Palo Alto,
  California, 1994.

\bibitem{cn2010}
R.~C{\'a}novas and G.~Navarro.
\newblock Practical compressed suffix trees.
\newblock In {\em Proc. SEA}, LNCS 6049, pages 94--105, 2010.

\bibitem{cl1994}
W.~I. Chang and E.~L. Lawler.
\newblock Sublinear approximate string matching and biological applications.
\newblock {\em Algorithmica}, 12(4--5):327--344, 1994.

\bibitem{cps2008}
G.~Chen, S.~J. Puglisi, and W.~F. Smyth.
\newblock {Lempel-Ziv} factorization using less time and space.
\newblock {\em Mathematics in Computer Science}, 1(4):605--623, 2008.

\bibitem{c1992}
M.~Crochemore.
\newblock String-matching on ordered alphabets.
\newblock {\em Theoretical Computer Science}, 92(1):33--47, 1992.

\bibitem{fgm2012}
P.~Ferragina, T.~Gagie, and G.~Manzini.
\newblock Lightweight data indexing and compression in external memory.
\newblock {\em Algorithmica}, 63(3):707--730, 2012.

\bibitem{fm2005}
P.~Ferragina and G.~Manzini.
\newblock Indexing compressed text.
\newblock {\em Journal of the ACM}, 52(4):552--581, 2005.

\bibitem{fh2007}
J.~Fischer and V.~Heun.
\newblock A new succinct representation of {RMQ}-information and improvements
  in the enhanced suffix array.
\newblock In {\em Proc. ESCAPE}, LNCS 4614, pages 459--470, 2007.

\bibitem{ggknp2012}
T.~Gagie, P.~Gawrychowski, J.~K{\"a}rkk{\"a}inen, Y.~Nekrich, and S.~J.
  Puglisi.
\newblock A faster grammar-based self-index.
\newblock In {\em Proc. LATA}, LNCS 7183, pages 240--251, 2012.

\bibitem{kkp2013}
J.~K{\"a}rkk{\"a}inen, D.~Kempa, and S.~J. Puglisi.
\newblock Linear time {L}empel--{Z}iv factorization: Simple, fast, small, 2012.
\newblock Manuscript, {\tt http://arxiv.org/abs/1212.2952}.

\bibitem{KarkkainenMP09}
J.~K{\"a}rkk{\"a}inen, G.~Manzini, and S.~J. Puglisi.
\newblock Permuted {L}ongest-{C}ommon-{P}refix array.
\newblock In {\em Proc. CPM}, LNCS 5577, pages 181--192, 2009.

\bibitem{klaap2001}
T.~Kasai, G.~Lee, H.~Arimura, S.~Arikawa, and K.~Park.
\newblock Linear-time longest-common-prefix computation in suffix arrays and
  its applications.
\newblock In {\em Proc. CPM}, LNCS 2089, pages 181--192, 2001.

\bibitem{kp2013}
D.~Kempa and S.~J. Puglisi.
\newblock {L}empel-{Z}iv factorization: fast, simple, practical.
\newblock In {\em Proc. ALENEX}, pages 103--112. SIAM, 2013.

\bibitem{kn2010}
S.~Kreft and G.~Navarro.
\newblock {LZ77}-like compression with fast random access.
\newblock In {\em Proc. DCC}, pages 239--248, 2010.

\bibitem{kn2011}
S.~Kreft and G.~Navarro.
\newblock Self-indexing based on {LZ77}.
\newblock In {\em Proc. CPM}, LNCS 6661, pages 41--54, 2011.

\bibitem{RLZspire2010}
S.~Kuruppu, S.~J. Puglisi, and J.~Zobel.
\newblock Relative {L}empel-{Z}iv compression of genomes for large-scale
  storage and retrieval.
\newblock In {\em Proc. SPIRE}, pages 201--206, 2010.

\bibitem{mm1993}
U.~Manber and G.~W. Myers.
\newblock Suffix arrays: a new method for on-line string searches.
\newblock {\em SIAM Journal on Computing}, 22(5):935--948, 1993.

\bibitem{nav2004}
G.~Navarro.
\newblock Indexing text using the {Z}iv-{L}empel trie.
\newblock {\em Journal of Discrete Algorithms}, 2(1):87--114, 2004.

\bibitem{n2012}
G.~Navarro.
\newblock Indexing highly repetitive collections.
\newblock In {\em Proc. IWOCA}, LNCS 7643, pages 274--279, 2012.

\bibitem{nm2007}
G.~Navarro and V.~M{\"a}kinen.
\newblock Compressed full-text indexes.
\newblock {\em ACM Computing Surveys}, 39(1):article 2, 2007.

\bibitem{og2011}
E.~Ohlebusch and S.~Gog.
\newblock {L}empel-{Z}iv factorization revisited.
\newblock In {\em Proc. CPM}, LNCS 6661, pages 15--26, 2011.

\bibitem{os2008}
D.~Okanohara and K.~Sadakane.
\newblock An online algorithm for finding the longest previous factors.
\newblock In {\em Proc. ESA}, LNCS 5193, pages 696--707, 2008.

\bibitem{os2009}
D.~Okanohara and K.~Sadakane.
\newblock A linear-time {B}urrows-{W}heeler transform using induced sorting.
\newblock In {\em Proc. SPIRE}, LNCS 5721, pages 696--707, 2009.

\bibitem{s2012}
T.~Starikovskaya.
\newblock Computing {L}empel-{Z}iv factorization online.
\newblock In {\em Proc. MFCS}, LNCS 7464, pages 789--799, 2012.

\bibitem{ZL77}
J.~Ziv and A.~Lempel.
\newblock A universal algorithm for sequential data compression.
\newblock {\em IEEE Transactions on Information Theory}, 23(3):337--343, 1977.

\end{thebibliography}

\end{document}